\documentclass[12pt]{article}
\usepackage{amsfonts}
\newcommand{\p}[1]{(\ref{#1})}
\newcommand{\cp}{\mbox{$\cal P$}}
\newcommand{\e}{\eta}
\newcommand{\be}{\begin{equation}}
\newcommand{\bea}{\begin{eqnarray}}
\newcommand{\ee}{\end{equation}}
\newcommand{\eea}{\end{eqnarray}}

\begin{document}
\topmargin-1cm
\renewcommand{\thefootnote}{\fnsymbol{footnote}}

\begin{center}
{\bf
Massless Higher Spin Fields
 in the AdS Background and BRST Constructions for Nonlinear Algebras
}\vspace{0.3cm} \\

I.L. Buchbinder${}^{a}$ \footnote{E-mail:
joseph@tspu.edu.ru}, A. Pashnev${}^{b}$\footnote{E-mail:
pashnev@thsun1.jinr.dubna.su}
and
M. Tsulaia${}^{b,c}$\footnote{E-mail:
tsulaia@thsun1.jinr.ru}\\
\vspace{0.3cm}

${}^a${\it Department Theoretical Physics,
Tomsk State Pedagogical University}\\
{\it Tomsk, 634041, Russia}\\
\vspace{0.3cm}

${}^b${\it Bogoliubov Laboratory of Theoretical Physics, JINR} \\
{\it Dubna, 141980, Russia}\\
~\\
${}^c${\it The Andronikashvili Institute of Physics, Georgian Academy
of Sciences,}\\
{\it Tbilisi, 380077, Georgia} \vspace{.5cm}\\

{\bf Abstract}
\end{center}
\begin{center}
{\begin{minipage}{4.2truein}
                 \footnotesize
                 \parindent=0pt

The detailed description of the method of the construction
of the nilpotent BRST charges for nonlinear algebras
of constraints appearing in the description of the massless higher spin fields
on the $AdS_D$ background is presented.
It is shown that the corresponding BRST charge is not
uniquely defined, but this ambiguity
has no impact on the physical content of the theory.

\end{minipage}}\end{center}
                 \vskip 2em \par

\renewcommand{\thefootnote}{\arabic{footnote}}
\setcounter{footnote}0\setcounter{equation}0

BRST -- BFV  method
\cite{BFV}
is a powerful tool for the
quantization of the wide class of the physical
systems, invariant with respect to the gauge symmetry groups.
The physical content of the theory is singled
 out by the BRST quantization condition
\begin{equation} \label{quant}
Q|Phys \rangle =0
\end{equation}
for the  nilpotent $Q^2 =0$ BRST charge.
Obviously due to the nilpotence property of the BRST charge
the BRST quantization condition is gauge
invariant i.e., the physical states
are defined up to the transformation
$|Phys \rangle' = |Phys \rangle + Q|\Lambda \rangle$
where the last term is a ``spurious" state
with the  zero norm. Therefore to find
the physical spectrum of the theory, one
 has to solve a cohomology problem of the corresponding
BRST charge.

On the other hand the BRST method
can be applied for the construction of the field theory, which corresponds
to some first quantized system.
Namely if one succeeds to construct the nilpotent
 BRST charge for the given system of constraints,
 then one can straightforwardly write the corresponding
 field theoretical Lagrangian of the form
$L \sim \langle \Phi|Q|\Phi \rangle$ which gives the condition
\p{quant} as the equation of motion. At the same time
the Lagrangian is gauge invariant under the transformations
\begin{equation} \label {tr}
\delta |\Phi \rangle = Q|\Lambda \rangle,
\end{equation}
The later property is crucial for eliminating   nonphysical
degrees of freedom from the spectrum.
The famous example of such kind of construction is BRST string field
theory.

The method of construction of the nilpotent BRST charge is well known
when the corresponding constraints form the closed linear Lie algebra.
The situation becomes more complicated when one deals with the second
-- class constraints and (or)
the algebra of constraints is nonlinear.
In the present paper we analyze this situation and show,
that
at least for the system of constraints describing irreducible
massless higher spin fields
 the corresponding BRST charge is not
unique. Rather it can contain a number of free parameters which, nevertheless
does not affect the physical content of the corresponding field theory.
These parameters, leading to the noncanonical form of the BRST charge, can be
present even in the case of closed linear algebra of constraints.
 The particular choice of these parameters
ensures the possibility to simplify the expression for the BRST charge,
and correspondingly the procedure of construction of  the field
theoretical Lagrangian becomes simpler as well.

Below we concentrate on the construction of the nilpotent BRST charges
and corresponding field theoretical Lagrangian for the theory describing the
higher spin fields in the flat space and in AdS spaces.
In particular, we investigate an arbitrariness in the
BRST charge, which is described by some free parameters, and show that
the physical content of the corresponding field theory is independent of
the values of these parameters.

First we briefly
describe the system of the free higher massless spin fields
propagating through the $D$ -- dimensional flat Minkowski space
time in the framework of the BRST approach \cite{PT2} and then generalize the method
for the interaction with $AdS_D$ background \cite{BuchPT}.

For the free massless higher spin fields the corresponding
system
of constraints imposed on the physical states includes
the mass -- shell constraint
$L_0 = p_\mu^2,$
transversality condition
$L_1 = p_\mu a^\mu,$
and tracelessness condition
$L_2= \frac{1}{2}a_\mu a^\mu,$
where
\be \label{osc}
p_\mu = \partial_\mu, \quad
\left[ a_\alpha,a_\beta^+ \right] =\eta_{\alpha \beta},\;\quad
\eta_{\alpha \beta}=diag(-1,1,1,...,1),
\ee
These constraints are written down in terms of creation and annihilation
operators $a_\mu^+, a^\mu$.
The general vector in the Fock space generated by these operators has the form
\be \label{Fockvector}
|\Phi\rangle =\sum_n
\Phi_{\alpha_1\alpha_2\cdots\alpha_{n}}
(x)
a^{\alpha_1 +}a^{\alpha_2 +}\cdots a^{\alpha_n +}
|0\rangle.\nonumber
\ee
Obviously, the high spin fields $\Phi_{\alpha_1\alpha_2\cdots\alpha_{n}}(x)$
are automatically symmetrical with respect to the permutation of their indices and therefore
representations of the corresponding flat space little group $O(d-2)$
are characterized by the Young tableaux with one row. Note, that more complicated Young tableaux
need for their description more then one set of creation and annihilation
operators.

The operators $L_0, L_1, L_2$ along with their conjugates $L_1^+$ and $L_2^+$
obey the following commutation relations.
\begin{eqnarray} \label{al1}
&&[L^+_1, L_1] = - L_0,  \quad
[L^+_1, L_2] = -L_1, \quad
[L^+_2, L_1] = -L^+_1, \nonumber \\
&&[G_0, L_1] = -L_1, \quad [L^+_1, G_0] = -L^+_1,
\end{eqnarray}
and
\begin{equation} \label{so21}
[G_0, L_2] = -2L_2, \quad
[L^+_2, G_0] = -2L^+_2, \quad
[L^+_2, L_2] = -G_0.
\end{equation}
where we have denoted $G_0 \equiv a^{+ \mu} a_\mu + \frac{D}{2}$.

The construction of the nilpotent BRST charge for this system of constraints
is not straightforward since as it can be seen from \p{so21}
the constraints $L_2^\pm$ are of the second class.
It is a consequence of the fact that the operator $G_0$ is strictly positive
and therefore can not be considered as an additional constraint.

Let us first introduce the anticommuting ghost variables
$\e_0, \e^+_1, \e_1, \e^+_2, \e_2, \e_G$ which correspond to the operators
$ L_0, L_1, L^+_1, L_2, L_2^+$ and $G_0$ and have the ghost number equal to $1$,
as well as the corresponding momenta
$\cp_0, \cp_1, \cp^+_1, \cp_2, \cp^+_2, \cp_G$ with the ghost number $-1$
obeying the anticommutation relations
\begin{equation}
\{\e_0,\cp_0\}=\{\e_2,\cp_2^+\}=\{\e_2^+,\cp_2\}=\{\e_2,\cp_2^+\}=\{\e_2^+,\cp_2\}=
\{\e_G,\cp_G\}=1.
\end{equation}
Then we build the auxiliary representations of the generators of the
algebra \p{so21} in terms of the additional creation and annihilation operators
$b^+$ and $b$, $[b, b^+]=1$ with the help of the following construction
using the Verma module.
We introduce the vector in the space of the Verma module
$|n\rangle_V = {(L_2^+)}^n|0\rangle_V$, $n \in N,$ $L_2|0\rangle_V=0$
and the corresponding vector in the Fock space
$$
|n \rangle = {(b^+)}^n|0\rangle, \quad b|0 \rangle =0.
$$
Mapping the vector in the space of Verma module into the vector in the
Fock space one obtains the representations of the generators of algebra
$SO(2,1)$ in terms of the variables $b^+$ and $b$
\begin{equation}
\label{aux} L^+_{2.aux} = b^+, \quad G_{0.aux} = 2b^+  b + h, \quad
L_{2.aux} = b^+  bb + bh
\end{equation}
where $h$ is the highest weight of the Verma module $G_{0.aux}|0 \rangle =h |0 \rangle$.
Then defining the operators $\tilde L^\pm_2 = L^\pm_2 + L^\pm_{2.aux},
\tilde G_0 = G_0 + G_{0.aux}$, we construct
the corresponding ``bare" nilpotent BRST charge for the
total system of constraints
\p{al1}--\p{so21}
\begin{eqnarray} \label {brst1}
\tilde Q \!\!& = &\!\!\!\e_0L_0+ \!\e_1 L^+_1 \! +
     \! \e^+_1  L_1\! +\!\e_2 \tilde L^+_2 + \!\!
\!\! \e^+_2\tilde L_2 \!\! +
     \nonumber \\
      &&\!\! +\e_G (-\tilde G_0  +3-
     \e^+_1  \cp_1  - \cp^+_1  \e_1-2 \cp^+_2  \e_2-2 \e^+_2   \cp_2) \nonumber \\
     &&\!\! +\e^+_2  \e_2
 \cp_G-
      \!\!  \e^+_1  \e_1  \cp + \e^+_1  \cp^+_1  \e_2-
     \e^+_2  \e_1  \cp_1.
\end{eqnarray}

In order to avoid the condition of the type $G_0| \Phi \rangle =0$
one has to get rid of the variables $\e_G$ and $\cp_G$ in the BRST charge
\p{brst1}, keeping its nilpotence property at the same time.
This can be  done by replacing the parameter $h$ by the expression
\begin{equation} \label{pi}
 \pi = -(G_0 + 2b^+b -3+
     \e^+_1  \cp_1  + \cp^+_1  \e_1+2 \cp^+_2  \e_2+2 \e^+_2   \cp_2)
\end{equation}
and then simply omit the dependence on  $\cp_G$ in the BRST
 charge (see \cite{BPT1} -- \cite{BPT2} for the general treatment).

 The ``reduced" BRST charge thus takes the form
\begin{eqnarray} \label{brst}
Q& =&\!\!\e_0 L_0 \!
  +\! \e_2(L^+_2 + b^+)\!
       +\!\e_1L^+_1\!  + \!\e^+_1  L_1\! +
    \!  \e^+_2 ( L_2- G_0b + b - b^+bb)\! \nonumber \\
       &&-
      \e^+_1  \e_1  \cp
      + \e^+_1  \cp^+_1  \e_2  -
      \e^+_2  \e_1  \cp_1 +
 \e^+_1  \e^+_2    \cp_1 b
       - \e^+_2  \cp^+_1    \e_1 b -
      2 \e^+_2  \cp^+_2    \e_2 b
\end{eqnarray}
Finally to restore the hermiticity property of the BRST charge, which
is lost due to the nonhermitian form of auxiliary representations \p{aux}
one has to define the scalar product in the Fock space as
$\langle \Phi_1|K|\Phi_2 \rangle$ were $K$ is a nondegenerate kernel operator
\be \label{yadro}
K= Z^+Z, \quad Z=\sum_{n} \frac{1}{n!}
 {(L_2^+)}^n |0 \rangle_V \langle 0|{(b)}^n
\ee
satisfying the property $KQ=Q^+K$.

Let us define the ghost vacuum in the way that the operators
$\e_0, \e_1^+, \e_2^+, \cp_1^+$ and $\cp_2^+$ are creation operators.
Then the BRST invariant Lagrangian for our system will be.
\begin{equation} \label{LF}
L= \int \e_0\langle \chi |KQ|\chi \rangle
\end{equation}
The state vector $|\chi \rangle$
along with the oscillators $a_\mu^+$ and $b^+$
depends on the ghost creation
operators as well and therefore the Lagrangian
\p{LF} contains some auxiliary fields.
However one can show that
after making  use of
the BRST gauge invariance \p{tr} and the
equations of motion with respect to the
 auxiliary fields, one is left with the only physical field
 with no ghost and $b^+$ dependence. This field
 is double traceless $L_2L_2|\Phi \rangle=0$ and the corresponding
 Lagrangian
 for this field has the form \cite{F}
\begin{equation} \label{LF1}
L=\langle \Phi|
 L_0 - L_1^+L_1 - 2 L^+_2  L_0  L_2
 +L^+_2  L_1  L_1
     +L^+_1  L^+_1  L_2   - L^+_2 L^+_1 L_1 L_2
|\Phi \rangle
\end{equation}
and the  Lagrangian \p{LF1}, is invariant under the gauge
transformations
\begin{equation} \label{GF}
\delta |\Phi \rangle
= L^+_1|\lambda_1 \rangle
\end{equation}
with the traceless parameter of gauge
transformations $L_2|\lambda_1 \rangle=0$.

Let us note that, the ``bare" nilpotent BRST charge \p{brst1}
has been constructed in the standard fashion adopted for the system of the
first -- class constraints i.e., the terms nonlinear in ghost variables
have the form given in \cite{BFV}. However there exists an ambiguity
in the expression of the BRST charge for this system
as we shall see below.

Now we generalize this procedure for the description
of the higher massless integer spin fields interacting with
$D$ -- dimensional Anti - de - Sitter background.
The main feature we face in this case is the nonlinear character
of the corresponding algebra of constraints \cite{BMV}.
In the AdS space the momentum operator $p_\mu$ (here we denote
Einstein indexes as $\mu, \nu, ...$
and the tangent space indexes as $\alpha, \beta, ...$)
 gets modified as
\begin{equation}
p_\mu = \partial_\mu + \omega_\mu{}^{\alpha \beta}a^+_\alpha a_\beta.
\end{equation}
Using this equation one cane easily check that the operators
$L_1^\pm = a^{\mu \pm}p_\mu$ are mutually hermitian conjugated with respect to
the integration measure $d^D x\sqrt {-g}$.
After introducing the covariant d'Alambertian
\begin{equation}
L_0 = g^{\mu \nu}(p_\mu p_\nu - \Gamma^\lambda_{\mu \nu}p_\lambda)
\end{equation}
one obtains  along with \p{so21} the following commutation relations
\begin{eqnarray} \label{al}
&&[L^+_1, L_1] = -\tilde L_0,  \quad
[L^+_1, L_2] = -L_1, \quad
[L^+_2, L_1] = -L^+_1, \nonumber \\
&&[\tilde L_0, L_1] = -2rL_1 + 4rG_0  L_1 -
      8rL^+_1  L_2, \nonumber\\
&&[L^+_1, \tilde L_0] = -2rL^+_1 + 4rL^+_1  G_0 -
      8rL^+_2  L_1,  \nonumber \\
&&[G_0, L_1] = -L_1, \quad [L^+_1, G_0] = -L^+_1,
\end{eqnarray}
where
\begin{equation} \label{ln}
\tilde L_0 \equiv L_0 + r(-D + \frac{D^2}{4}) + 4rL^+_2  L_2 - rG_0
G_0 + 2rG_0,
\end{equation}
and the parameter $r$ being related to the inverse radius of the
$AdS_D$ space $\lambda$ via $r = \lambda^2$.

As it can be seen from the relations \p{al} and \p{so21}
the operators obey the nonlinear algebra being analogous to the finite
$W_3^{(2)}$ algebra \cite{dBHT}. Again our goal is
 to construct the nilpotent BRST charge
for this system, where the
operator $G_0$ is excluded from the total set of constraints.

As a first step we construct nilpotent BRST charge
for the operators $L_0, L_1^\pm, L_2^\pm$ and $ G_0$ satisfying the
nonlinear algebra \p{so21}, \p{al}.
However in contrast
to the case of the flat space -- time
i.e.,
when the constraints
form the linear algebra
there is no standard prescription for the construction of nilpotent BRST charge
for the system of operators  forming a nonlinear algebra.
For this end we add
step by step all possible
higher order terms in the ghosts with arbitrary coefficients to the BRST charge \p{brst1}
and require its nilpotence.
Simultaneously we modify the operator at the linear term in ghost variable $\e_0$,
instead of being simply $L_0$ it becomes
the nonlinear combination $\tilde L_0$ of $L_0$, $G_0$ and $L_2^\pm$ \p{ln}.

This procedure leads to
the family of ``bare" nilpotent
 BRST charges, which  turn out to depend on
three free parameters $k_1,k_2,k_3,$ namely i.e, the
expression of  terms which contain
higher degrees in the ghost variables turn out  not to be uniquely defined.
The possible solution of the problem (not necessarily the most general one)
has the form
\begin{equation}\label{family}
  \tilde Q^1=\tilde Q^1_0+k_1 \tilde Q^1_{k_1}+
   k_2 \tilde Q^1_{k_2}+k_3 \tilde Q^1_{k_3},
\end{equation}
where
\begin{eqnarray}
\tilde Q^1_0 &=&  \e_0 (\tilde L_0 + 6r)
+ \e_1L_1^+ +
    \e_2L_2^+ + \e_1^+L_1 + \e_2^+L_2 - \e_G(G_0 -3) \nonumber \\
        &&  +2r \e_0\e_1^+\cp_1
        -8r
     \e_0\e_2^+\cp_2
      + 2r \e_0\cp_1^+\e_1 \nonumber \\
    &&-8r \e_0\cp_2^+\e_2 -
    \e_1^+\e_1\cp_0
    + \e_1^+\e_G\cp_1 + \e_1^+\cp_1^+\e_2
    -4r \e_1^+\cp_G\e_1 \nonumber \\
    &&- \e_2^+\e_1\cp_1
    +2\e_2^+\e_G\cp_2 - \e_2^+\cp_G\e_2 +
    \cp_1^+\e_G\e_1 + 2\cp_2^+\e_G\e_2
-8r \e_0\e_1\cp_1L_2^+ \nonumber \\
    &&-4r \e_0\e_1^+\cp_1G_0
     + 8r
     \e_0\e_1^+\cp_1^+L_2
-4r \e_0\cp_1^+\e_1G_0
-12r
\e_0\e_1^+\cp_1^+\e_1\cp_1
\\
\tilde Q^1_{k_1} &=&  -2\e_0(G_0 - 3)  - 6 \e_0\e_2^+\cp_2
- 6\e_0\cp_2^+\e_2
- 3\e_1^+\cp_G\e_1
     - 2\e_0\e_1\cp_1L_2^+ \nonumber \\
&&- \e_0\e_1^+\cp_1G_0
    -2
     \e_0\e_1^+\cp_2L_1^+ + 2
     \e_0\e_1^+\cp_1^+L_2
      - \e_0\e_1^+\cp_GL_1  \nonumber \\
     &&-
      \e_0\cp_1^+\e_1G_0
    -2
     \e_0\cp_2^+\e_1L_1
     - \e_0\cp_G\e_1L_1^+
 - 6
     \e_0\e_1^+\cp_1^+\e_1\cp_1,
\\
\tilde Q^1_{k_2}&=&-\e_0(G_0 -3)
- \e_0\e_1^+\cp_1
     -2\e_0\e_2^+\cp_2
       - \e_0\cp_1^+\e_1
     - 2\e_0\cp_2^+\e_2
-  \e_1^+\cp_G\e_1
\\
\tilde Q^1_{k_3}&=&\e_0\cp_2^+\e_2 +\e_0\e_2^+\cp_2 + \e_0\e_1\cp_1L_2^+
+ \e_0\e_1^+\cp_2L_1^+
-\e_0\e_1^+\cp_1^+L_2
+\e_0\cp_2^+\e_1L_1 \nonumber \\
&&  +
    \e_0\e_1^+\e_2^+\cp_1\cp_2 +
    2 \e_0\e_1^+\cp_1^+\e_1\cp_1
+\e_0\e_1^+\cp_1^+\cp_G\e_2 -
    \e_0\e_1^+\cp_2^+\e_2\cp_1 \nonumber \\
     &&-
    \e_0\e_2^+\cp_1^+\e_1\cp_2 -
    \e_0\e_2^+\cp_G\e_1\cp_1 +
    \e_0\cp_1^+\cp_2^+\e_1\e_2
    \end{eqnarray}

All these operators are nilpotent and mutually anticommuting so that their
sum is  nilpotent as well.
Let us note  that the particular choice of the parameters
$ k_1=0, k_2=0, k_3=8$
leads to the BRST charge constructed in \cite{BuchPT}, which includes
terms up to the fifth order in ghosts.
The  simplest expression including
only third powers of ghosts corresponds to the choice $ k_1=-2, k_2=4, k_3=0 $:
\begin{eqnarray} \label{simp}
\tilde Q^1& =& \e_0( \tilde L_0 + 6r)
  +
    \e_1L_1^+ + \e_2L_2^+
    + \e_1^+L_1 + \e_2^+L_2 \nonumber \\
    &&-
    \e_G(G_0-3)
-
    2r\e_0\e_1^+\cp_1 - 4r\e_0\e_2^+\cp_2 \nonumber \\
    &&- 2r\e_0\cp_1^+\e_1 -
    4r\e_0\cp_2^+\e_2 - \e_1^+\e_1\cp_0 + \e_1^+\e_G\cp_1
    +
    \e_1^+\cp_1^+\e_2 - 2r\e_1^+\cp_G\e_1 \nonumber \\
    && -
    \e_2^+\e_1\cp_1 + 2\e_2^+\e_G\cp_2 -
    \e_2^+\cp_G\e_2 + \cp_1^+\e_G\e_1 +
    2\cp_2^+\e_G\e_2 - 4r\e_0\e_1\cp_1L_2^+ \nonumber \\
    && -
    2r\e_0\e_1^+\cp_1G_0 + 4r\e_0\e_1^+\cp_2L_1^+ +
    4r\e_0\e_1^+\cp_1^+L_2
    + 2r\e_0\e_1^+\cp_GL_1 \nonumber \\
    &&-
    2r\e_0\cp_1^+\e_1G_0 + 4r\e_0\cp_2^+\e_1L_1
    +
    2r\e_0\cp_G\e_1L_1^+ .
\end{eqnarray}

To get rid of the nonphysical constraint $G_0$ we again
take the auxiliary representations for the operators
$L_2^\pm$ and $G_0$ in the form \p{aux} and
construct the  ``bare" nilpotent BRST charge
adding step by step all possible
higher order terms in the ghosts. The result is
similar to  \p{family}, namely
\begin{equation}\label{family1}
  \tilde Q=\tilde Q_0 + k_1 \tilde Q_{k_1}+ k_2 \tilde Q_{k_2}+
  k_3 \tilde Q_{k_3},
\end{equation}
but now
\begin{eqnarray}
\tilde Q_0& =& \e_0(\tilde L_0 +6r
    -4r G_{0.aux})
  +
     \e_1L_1^+
    +\e_1^+L_1 \nonumber \\
    && + \e_2(L^+_{2.aux} + L_2^+)+ \e_2^+(L_{2.aux} + L_2) - \e_G(G_{0.aux} +
    G_0 -3) \nonumber \\
      &&+2r \e_0\e_1^+\cp_1
-8r
     \e_0\e_2^+\cp_2 + 2r \e_0\cp_1^+\e_1 \nonumber \\
     &&
    -8r \e_0\cp_2^+\e_2 -
    \e_1^+\e_1\cp_0 + \e_1^+\e_G\cp_1 + \e_1^+\cp_1^+\e_2
-4r \e_1^+\cp_G\e_1 - \e_2^+\e_1\cp_1 \nonumber \\
&&+
    2\e_2^+\e_G\cp_2 - \e_2^+\cp_G\e_2 +
    \cp_1^+\e_G\e_1
    + 2\cp_2^+\e_G\e_2
-8r \e_0\e_1\cp_1L_2^+ \nonumber \\
    &&
-4r \e_0\e_1^+\cp_1G_0 +
    8r \e_0\e_1^+\cp_1^+L_2 -
4r \e_0\cp_1^+\e_1G_0
-12r
     \e_0\e_1^+\cp_1^+\e_1\cp_1,
\\
\tilde Q_{k_1}& =&    -2\e_0(G_{0.aux} + G_0 -3)
- 6 \e_0\e_2^+\cp_2
- 6\e_0\cp_2^+\e_2
- 3\e_1^+\cp_G\e_1  \nonumber \\
&&-2 \e_0\e_1\cp_1(L^+_{2.aux} +
     L_2^+)
     +
     2 \e_0\e_1^+\cp_1^+(L_{2.aux} +L_2)
    -
    \e_0\e_1^+\cp_1(G_{0.aux}  + G_0) \nonumber \\
    &&
    -2\e_0\e_1^+\cp_2L_1^+
      -
    \e_0\e_1^+\cp_GL_1
    - \e_0\cp_1^+\e_1(G_{0.aux} +G_0) \nonumber \\
     && -2\e_0\cp_2^+\e_1L_1
     - \e_0\cp_G\e_1L_1^+
- 6
     \e_0\e_1^+\cp_1^+\e_1\cp_1,
\\
\tilde Q_{k_2} &=&- \e_0(  G_{0.aux}
      +  G_0 -3)
- \e_0\e_1^+\cp_1
- 2
     \e_0\e_2^+\cp_2  \nonumber \\
&&- \e_0\cp_1^+\e_1
      - 2\e_0\cp_2^+\e_2 -
\e_1^+\cp_G\e_1,
\\
\tilde Q_{k_3}& =&\e_0\e_2^+\cp_2 +
    \e_0\cp_2^+\e_2
+
    \e_0\e_1\cp_1(L^+_{2.aux} + L_2^+) -
    \e_0\e_1^+\cp_1^+(L_{2.aux} + L_2) \nonumber \\
    &&+\e_0\e_1^+\cp_2L_1^+
+
     \e_0\cp_2^+\e_1L_1
+\e_0\e_1^+\e_2^+\cp_1\cp_2 + 2
     \e_0\e_1^+\cp_1^+\e_1\cp_1 \nonumber \\
     &&+
    \e_0\e_1^+\cp_1^+\cp_G\e_2 -
    \e_0\e_1^+\cp_2^+\e_2\cp_1
    -
    \e_0\e_2^+\cp_1^+\e_1\cp_2 \nonumber \\
    &&-
    \e_0\e_2^+\cp_G\e_1\cp_1 +
    \e_0\cp_1^+\cp_2^+\e_1\e_2.
\end{eqnarray}
Taking the free parameters as in
\p{simp} again leads to the simplest form of
the BRST charge
\begin{eqnarray} \label{tilde}
\tilde Q &=& \e_0(\tilde L_0 + 6r - 4rG_{0.aux})
 + \e_1L_1^+ +
    \e_1^+L_1 \nonumber \\
    && + \e_2(L^+_{2.aux} + L_2^+)
    + \e_2^+(L_{2.aux} + \e_2^+L_2) - \e_G(G_{0.aux} +
    G_0 -3) \nonumber \\
    && -
    2r\e_0\e_1^+\cp_1 - 4r\e_0\e_2^+\cp_2
     - 2r\e_0\cp_1^+\e_1
     -
    4r\e_0\cp_2^+\e_2 - \e_1^+\e_1\cp_0 \nonumber \\
    &&+ \e_1^+\e_G\cp_1
      +
    \e_1^+\cp_1^+\e_2 - 2r\e_1^+\cp_G\e_1
    -
    \e_2^+\e_1\cp_1 \nonumber \\
    &&+ 2\e_2^+\e_G\cp_2
    -
    \e_2^+\cp_G\e_2 + \cp_1^+\e_G\e_1 +
    2\cp_2^+\e_G\e_2 + 4r\e_0\e_1\cp_1L^+_{2.aux} \nonumber \\
    && -
    4r\e_0\e_1\cp_1L_2^+ + 2r\e_0\e_1^+\cp_1G_{0.aux} -
    2r\e_0\e_1^+\cp_1G_0 + 4r\e_0\e_1^+\cp_2L_1^+ \nonumber \\
    && -
    4r\e_0\e_1^+\cp_1^+L_{2.aux} + 4r\e_0\e_1^+\cp_1^+L_2 +
    2r\e_0\e_1^+\cp_GL_1 + 2r\e_0\cp_1^+\e_1G_{0.aux} \nonumber \\
    &&-
    2r\e_0\cp_1^+\e_1G_0 + 4r\e_0\cp_2^+\e_1L_1 +
    2r\e_0\cp_G\e_1L_1^+.
\end{eqnarray}

The further procedure goes on in the complete analogy with the case of the
flat space -- time background. Namely we obtain the reduced BRST charges $Q$
from \p{family1} after the transformation \p{pi} and than construct the
 Lagrangian \p{LF} being invariant under the gauge transformation
\p{tr}.
Though the BRST charge which corresponds to the system \p{al}, \p{so21}
is not unique, one can show following the lines
of \cite{BuchPT} on the base of the explicit calculations
 that all these BRST charges after making the partial BRST gauge fixing
in the Lagrangian \p{LF} lead to the unique final form of the Lagrangian,
which contains only one double traceless physical field $|\Phi \rangle$
($L_2L_2|\Phi \rangle=0$)
\begin{eqnarray} \label{ALF}
\cal L&=&\langle \Phi|
 \tilde L_0 - L_1^+L_1 - 2 L^+_2  \tilde L_0  L_2
 +L^+_2  L_1  L_1
     +L^+_1  L^+_1  L_2   - L^+_2 L^+_1 L_1 L_2 \nonumber \\
 &&- r(6  - 4    G_0 + 10    L^+_2  L_2      -
    4  L^+_2  G_0  L_2)
|\Phi \rangle.
\end{eqnarray}
The Lagrangian is invariant with respect to the gauge transformation \p{GF}
with the traceless parameter $|\lambda \rangle$ and describes
massless irreducible higher spins in AdS space which correspond to the Young
tableaux with one row in complete correspondence with \cite{BuchPT}.
The same result concerning non-uniqueness of the BRST charge will take
place if we apply above construction to the system of constraints
\p{al1} -- \p{so21} describing the propagation of higher massless
integer spin fields in the Minkowski space as well.

\setcounter{equation}0\section{Conclusions}
In this paper we have shown that the BRST charge which corresponds
to the system of constraints describing higher massless integer spin fields
is not unique. Rather there exists a family of charges even for the case
of the flat space background. As a result the form and the number of
terms which contain the third and higher degrees of the
ghost variables
may be different. However the final form
of the corresponding gauge invariant
Lagrangian and of the gauge transformations are such,
that the  physical spectrum of the theory
does not depend on the particular choice of
free parameters entering into the BRST charge.
That means that in this particular case the cohomologies
of the all family BRST charges given by the equation
\p{family1} both for the case of the flat and $AdS_D$ backgrounds
are the same.
Moreover we conjecture that generically
 the analogous ambiguity if it is observed for  other
gauge systems should have no impact on the physical  spectrum
for the different particular choices of the free parameters. The number of free parameters
can be different as well.
Let us note also that the nilpotence of
the BRST charge for the case of $AdS_D$ background requires
the modification of the mass -- shell operator,
which leads the  constraint, giving the correct
unitary massless representations of $AdS_D$ group.

It seems interesting to apply this procedure to the more complicated representations
of the $AdS_D$ group, when the corresponding representations
of the flat space time little group
is described for two and more rows.
It is interesting also to construct the auxiliary representations
of the whole nonlinear algebra \p{al}, \p{so21}
in
terms of Verma module, as well
as to study the possibility for the construction of finite
dimensional representations for such kind of algebras.

\noindent {\bf Acknowledgments}
The work of the authors was supported in part by
the INTAS grant, project No 00-00254 and by the RFBR grants.
The work of I.L.B and A.P
was also supported in part by the joint DFG-RFBR grant.
 I.L.B is grateful to the INTAS grant, project No 991-590 for partial
support. The work of A.P and M.T was supported in part by
 Votruba-Blokhintsev (Czech Republic-BLTP JINR) grant.

\end{document}